\documentclass[american,english,aip,preprint]{revtex4-1}
\usepackage[latin9]{inputenc}
\usepackage{amsmath}
\usepackage{amssymb}
\usepackage{graphicx}
\usepackage{esint}

\makeatletter

\usepackage{amsfonts}\usepackage{comment}

\usepackage{default}
\usepackage{babel}

\usepackage{babel}

\makeatother

\usepackage{babel}
\begin{document}

\title{Geodesic acoustic modes with poloidal mode couplings ad infinitum}

\author{Rameswar Singh$^{1}$, Ö D Gürcan$^{1}$, X Garbet$^{2}$, P Hennequin$^{1}$,
L Vermare$^{1}$, P Morel$^{1}$ and R Singh$^{3}$}

\affiliation{$^{1}$Laboratoire de Physique des Plasmas, Ecole Polytechnique,
91128 Palaiseau Cedex, France\\
$^{2}$CEA,IRFM, F-13108 Saint Paul Lez Durance, France\\
$^{3}$WCI, NFRI, Daejeon, Republic of Korea }

\email{rameswar.singh@lpp.polytechnique.fr}

\date{\today}
\selectlanguage{american}%
\begin{abstract}
Geodesic acoustic modes (GAMs) are studied, for the first time, including
all poloidal mode $(m)$ couplings using drift reduced fluid equations.
The nearest neighbor coupling pattern, due to geodesic curvature,
leads to a semi-infinite chain model of the GAM with the mode-mode
coupling matrix elements proportional to the radial wave number $k_{r}$.
The infinite chain can be reduced to a renormalized bi-nodal chain
with a matrix continued fractions. Convergence study of linear GAM
dispersion with respect to $k_{r}$ and the $m$-spectra confirms
that high $m$ couplings become increasingly important with $k_{r}$.
The radially sorted roots overlap with experimentally measured GAM
frequency profile in low collisionality shots in Tore Supra thus explaining
the reduced frequency of GAM in Tore Supra. 
\end{abstract}


\maketitle
\selectlanguage{english}%

\section{Introduction}

\selectlanguage{american}%
\noindent Control of turbulent transport is one of the key requirements
for the improvement of the confinement time necessary for achieving
controlled thermonuclear fusion. Large scale flow structures such
as zonal flows\cite{zlin:1998} that are self-generated by, and that
suppress the underlying micro turbulence and hence the turbulent transport
in tokamaks by shearing the turbulent eddies\cite{diamond:05}, provide
an important potentiality to this end. Geodesic acoustic modes (or
GAMs), first predicted by Winsor \emph{et al} \cite{winsor:68}, are
oscillating zonal flows, which also have a regulating effect on turbulence.
While zonal flows are mostly stationary, GAM oscillations are sustained
by geodesic curvature of the equilibrium magnetic field, which also
allows them to be observed\cite{jakubowski:2002,nagashima:2005,kramer-flecken:2006,zhao:2006,vermare:2012}.
GAMs are linearly damped via Landau damping \cite{sugama:2006}, and
thus exist mainly near the edge region, in a standard tokamak. They
may be excited by non-linear Reynolds stresses\cite{itoh:2005_2,nikhil:2007,nikhil:2008,guzdar:2008,zonca:epl2008,guzdar:2009,chen:2014},
poloidally asymmetric particle flux\cite{itoh:2005_2} and heat fluxes\cite{hallatschek:2001}
and linearly by energetic particles\cite{fu:2008,fu:2011,qui:2010,qui:2012,zarzoso:2012,wang:2014,girardo:2014}.
\\
Despite many progress in understanding of GAMs and their interaction
with the underlying turbulence, even the linear structure of the GAM
is still not perfectly understood. It is usually argued that the electrostatic
GAM is made of toroidal mode $n=0$, and poloidal mode number $m=0$
of potential $\delta\phi$ and $n=0$, $m=1$ of density $\delta n$,
temperature $\delta T$ and parallel velocity $\delta v_{\parallel}$
perturbations with $\sin\theta$ parity of $\delta n,\delta T$ and
$\cos\theta$ parity of $\delta v_{\parallel}$ in the poloidal direction.
But it is well known in the study of toroidal drift waves that the
magnetic curvature couples the poloidal mode numbers to their nearest
neighbors making $m$ a bad quantum number\cite{horton:78}. GAM is
no exception to this. These couplings are dropped usually beyond $m=1$
or sometimes at $m=2$\cite{nguyen:2008,zonca:epl2008}.\\
This paper revisits the theory of GAM with poloidal mode couplings
\emph{ad-infinitum. }For simplicity and to bring out the novel effects,
linear theory of GAMs are studied in a simple Ion temperature gradient
driven (ITG) turbulence with adiabatic electron response using the
drift reduced Braginskii equations\cite{zeiler:1997,braginskii:65},
which may be suitable for the edge region of a modern day tokamak.
The toroidal coupling of different poloidal modes that constitute
the GAM appear as a nearest neighbor coupling of an infinite set of
equations with appropriate parities in $\theta$. A vector recurrence
relation with appropriately defined orbitals leads to an elegant semi-infinite
chain model of GAM, which can be reduced to a renormalized bi-nodal
chain using a matrix continued fraction (MCF) approach\cite{risken:89}.
GAM dispersion at different chain lengths shows that the high $m$
couplings become more important with increasing $k_{r}$. Recently
it was observed that the GAM frequency in Tore Supra is $50\%$ smaller
than the theoretical predictions from gyrokinetic model\cite{storelli:2015}
and fluid models with effects of finite beta and collisionality and
keeping only the poloidal mode numbers up to $m=1$\cite{rameswar:ppcf2015}.
Our theoretical model is compared with the experimental GAM frequencies
observed in Tore Supra shots and it is found that the experimental
values of frequencies overlap with the radially sorted theoretical
values of GAM frequencies when high poloidal mode couplings are retained.\\
The rest of the paper is organized as follows. In Section\ref{sec:Drift-reduced-braginskii}
the drift reduced Braginskii equations are briefly discussed. Fully
nonlinear electrostatic GAM equations are derived from drift reduced
Braginskii equations in the Section\ref{sec:Electrostatic-collisionless-Geod}.
Experimenal comparisons are presented in Section\ref{sec:Numerical-solutions-and}
and the conclusions are drawn in Section\ref{sec:Conclusions}. 

\selectlanguage{english}%

\section{Drift reduced braginskii equations\label{sec:Drift-reduced-braginskii}}

\noindent We start with the electrostatic subset of the drift reduced
fluid equations derived in Refs\cite{zeiler:1997,rameswar:ppcf2015}
which were obtained from Braginskii equations\cite{braginskii:65}.
The space time scales are normalized as $r=r/\rho_{s}$, $\nabla_{\parallel}\equiv L_{n}\nabla_{\parallel}$,
$t=tc_{s}/{L_{n}}$. The \foreignlanguage{american}{perturbed} field
quantities are normalized to their mixing length levels: $\phi=({e\delta\phi}/{T_{e0}})({L_{n}}/{\rho_{s}})$,
$n_{i}=({\delta n_{i}}/n_{0})({L_{n}}/{\rho_{s}})$, $v=({\delta v_{\parallel i}}/{c_{s}})({L_{n}}/{\rho_{s}})$,
$p_{i}=({\delta p_{i}}/{P_{e0}})({L_{n}}/{\rho_{s}})$. The remaining
dimensionless parameters are : $\eta_{i}=L_{n_{i0}}/L_{T_{i0}}$,
$\varepsilon_{n}=2L_{n}/R$, $K=\tau_{i}(1+\eta_{i})$, $\tau_{i}=T_{i0}/T_{e0}$,
$\rho_{s}=c_{s}/\omega_{ci}$ is ion Larmor radius, and $c_{s}=\sqrt{T_{e0}/m_{i}}$
is ion sound speed where $m_{i}$ is ion mass. The nonlinearities
in the following equations originates mainly from $E\times B$ drift
nonlinearity i.e., $\vec{v}_{E\times B}\cdot\vec{\nabla}f=\left[\phi,f\right]$,
and polarization drift nonlinearity $\vec{v}_{E\times B}\cdot\vec{\nabla}\nabla_{\perp}^{2}f=\left[\phi,\nabla_{\perp}^{2}f\right]$
.

\subsection{Electron response}

In the collisionless limit the electron temperature perturbations
are washed out $T_{e}=0$. Then assuming the limit $\omega<<k_{||}c_{s}$
the parallel electron momentum equation yields the adiabatic electron
response 
\begin{eqnarray}
n_{e}=\phi
\end{eqnarray}

\subsection{Ion response}

The perturbed ion density equation is given by 
\begin{eqnarray}
 &  & \frac{\partial n_{i}}{\partial t}+\frac{1}{r}\frac{\partial\phi}{\partial\theta}-\varepsilon_{n}\left(\cos\theta\frac{1}{r}\frac{\partial}{\partial\theta}+\sin\theta\frac{\partial}{\partial r}\right)\left(\phi+\tau_{i}n_{i}+\tau_{i}T_{i}\right)-\left(\frac{\partial}{\partial t}-K\frac{1}{r}\frac{\partial}{\partial\theta}\right)\nabla_{\perp}^{2}\phi\nonumber \\
 &  & +\nabla_{||}v_{||}=-\left[\phi,n_{i}\right]+\vec{\nabla}\cdot\left[\phi+p_{i},\vec{\nabla}_{\perp}\phi\right]\label{ec}
\end{eqnarray}
Here the second term on the left hand side is the $E\times B$ convection
of equilibrium density, the third term comes from divergence of the
$E\times B$ velocity plus the divergence of diamagnetic particle
flux, the fourth term comes from the divergence of the polarization
drift and the fifth term simply represents the parallel compression.
On the right hand side, the first term is the $E\times B$ nonlinearity
and the second term is polarization nonlinearity. The Poisson brackets
are defined as $[a,b]=\frac{\partial a}{\partial r}\frac{1}{r}\frac{\partial b}{\partial\theta}-\frac{1}{r}\frac{\partial a}{\partial\theta}\frac{\partial b}{\partial r}$.
Here $r$, $\theta$, $\varphi$ represent the radial, poloidal and
toroidal coordinates in tokamak and circular flux surfaces are assumed
for the equilibrium. The equilibrium magnetic field is given by $\vec{B}=B_{\varphi}(\vec{e}_{\varphi}+\frac{r}{qR}\vec{e}_{\theta})$
where the toroidal magnetic field is $B_{\varphi}=\frac{B_{0}}{1+(r/R_{0})\cos\theta}$.
Adding the electron momentum equation to the ion momentum equation
and then using the electron temperature equation, the parallel ion
velocity equation becomes: 
\begin{eqnarray}
 &  & \frac{\partial v_{\parallel}}{\partial t}-2\tau_{i}\varepsilon_{n}\left(\cos\theta\frac{1}{r}\frac{\partial}{\partial\theta}+\sin\theta\frac{\partial}{\partial r}\right)v_{||}+\nabla_{||}\left[\tau_{i}n_{i}+\tau_{i}T_{i}+\phi\right]\nonumber \\
 &  & =-\left[\phi,v_{||}\right]+n_{i}\nabla_{||}p_{i}\label{iv}
\end{eqnarray}
The second term on the left hand side originates in parts from the
divergence of ion stress tensor after the gyroviscous cancellation.
The third term represents the sum of parallel pressure and electric
force. The first term on the right hand side is the $E\times B$ nonlinearity
and second term is the parallel pressure force nonlinearity. The ion
temperature equation reads 
\begin{eqnarray}
 &  & \frac{\partial}{\partial t}\left(T_{i}-\frac{2}{3}n_{i}\right)-\frac{5}{3}\varepsilon_{n}\left(\cos\theta\frac{1}{r}\frac{\partial}{\partial\theta}+\sin\theta\frac{\partial}{\partial r}\right)T_{i}+\left(\eta_{i}-\frac{2}{3}\right)\frac{1}{r}\frac{\partial\phi}{\partial\theta}\nonumber \\
 &  & =-\sqrt{2\tau_{i}}|\nabla_{\parallel}|T_{i}-\left[\phi,T_{i}-\frac{2}{3}n_{i}\right]
\end{eqnarray}
The second term on the left hand side originates from the divergence
of diamagnetic heat flux after cancellation in parts with diamagnetic
convection of heat. The first term on the right hand side represents
Landau damping \foreignlanguage{american}{à la} Hammet and Perkins\foreignlanguage{american}{\cite{hammet:1990}}.
The second term on the right represents nonlinear $E\times B$ convection
of heat. Finally the quasineutrality $n_{e}=n_{i}$ is used for the
perturbations in the form of the vorticity equation 
\begin{eqnarray}
 &  & \left(\frac{\partial}{\partial t}-K\frac{1}{r}\frac{\partial}{\partial\theta}\right)\nabla_{\perp}^{2}\phi+\varepsilon_{n}\left(\cos\theta\frac{1}{r}\frac{\partial}{\partial\theta}+\sin\theta\frac{\partial}{\partial r}\right)\left[\tau_{i}n_{i}+\tau_{i}T_{i}+n_{e}\right]\nonumber \\
 &  & =-\vec{\nabla}\cdot\left[\phi+p_{i},\vec{\nabla}_{\perp}\phi\right]
\end{eqnarray}
The first term on the left hand side is the divergence of polarization
current and the second term is the divergence of diamagnetic current.
The term on the right hand side is the polarization nonlinearity.

\section{Electrostatic collisionless Geodesic acoustic modes\label{sec:Electrostatic-collisionless-Geod}}

In the minimal description, the GAM is said to be made of poloidally
smooth potential and $\sin\theta$ parity of density perturbations.
However here we wish to derive the equations for GAM with all poloidal
mode couplings. Hence we start by taking the flux surface average
of the vorticity equation.

\begin{eqnarray}
\frac{\partial}{\partial t}\nabla_{r}^{2}\left\langle \phi\right\rangle +\varepsilon_{n}\nabla_{r}\left[\left(1+\tau_{i}\right)\left\langle n\sin\theta\right\rangle +\tau_{i}\left\langle T_{i}\sin\theta\right\rangle \right]=-\left\langle \vec{\nabla}\cdot\left[\phi+p_{i},\vec{\nabla}_{\perp}\phi\right]\right\rangle 
\end{eqnarray}
The angular bracket represents flux surface averaging defined by $\left\langle (...)\right\rangle =\int rd\theta d\varphi(...)/\int rd\theta d\varphi$.
The nonlinear term on the right hand side represents the divergence
of vorticity flux and is attributed to turbulent excitation of GAM\cite{nikhil:2007}.
This shows the need for an equation for $\left\langle n\sin\theta\right\rangle $
which is obtained by multiplying the ion continiuity equation by $\sin\theta$
followed by flux surface average. In the following a general equation
for poloidal mode number $m$ for $\left\langle n\sin m\theta\right\rangle $
is obtained. 
\begin{eqnarray}
 &  & \frac{\partial}{\partial t}\left\langle n\sin m\theta\right\rangle -\frac{\partial}{\partial t}\nabla_{r}^{2}\left\langle \phi\sin m\theta\right\rangle -\frac{\varepsilon_{n}}{2}\nabla_{r}\left[\left\langle \phi\cos(m-1)\theta\right\rangle +\tau_{i}\left\langle n\cos(m-1)\theta\right\rangle \right.\nonumber \\
 &  & \left.+\tau_{i}\left\langle T_{i}\cos(m-1)\theta\right\rangle -\left\langle \phi\cos(m+1)\theta\right\rangle -\tau_{i}\left\langle n\cos(m+1)\theta\right\rangle -\tau_{i}\left\langle T_{i}\cos(m+1)\theta\right\rangle \right]\nonumber \\
 &  & -\frac{\varepsilon_{n}}{2q}m\left\langle v_{||}\cos m\theta\right\rangle =-\left\langle \vec{\nabla}\cdot\left[\phi+p_{i},\vec{\nabla}_{\perp}\phi\right]\sin m\theta\right\rangle +\left\langle S_{n}\sin m\theta\right\rangle \label{zcs}
\end{eqnarray}
The first term on the right hand side is the $\sin m\theta$ weighted
flux surface averaged divergence of vorticity flux. The second term
represents $\theta$ antisymmetric part of the particle source $S_{n}$.
This then shows the need for the equation for $\left\langle v_{||}\cos m\theta\right\rangle $,
which is obtained by multiplying the parallel ion velocity equation
(\ref{iv}) by $\cos m\theta$ followed by flux surface average 
\begin{eqnarray}
 &  & \frac{\partial}{\partial t}\left\langle v_{||}\cos m\theta\right\rangle -\tau_{i}\varepsilon_{n}\nabla_{r}\left[\left\langle v_{\parallel}\sin(m+1)\theta\right\rangle -\left\langle v_{\parallel}\sin(m-1)\theta\right\rangle \right]\nonumber \\
 &  & +\frac{\varepsilon_{n}}{2q}m\left[\left(1+\tau_{i}\right)\left\langle n\sin m\theta\right\rangle +\tau_{i}\left\langle T_{i}\sin m\theta\right\rangle \right]=-\left\langle \left[\phi,v_{||}\right]\cos m\theta\right\rangle \nonumber \\
 &  & +\left\langle \left(n\nabla_{||}\left(p_{i}+n\right)\right)\cos m\theta\right\rangle +\left\langle S_{v}\cos m\theta\right\rangle \label{gv}
\end{eqnarray}
The various nonlinear terms on the right hand side of the above equation
can be identified as follows: The first term coming from the $E\times B$
convective nonlinearity is the $\cos m\theta$ weighted, flux surface
averaged, divergence of the parallel velocity flux. The second term
is the flux surface average of the turbulent parallel acceleration
weighted by $\cos m\theta$. This term survives only when there is
a $k_{\parallel}$ symmetry breaking mechanism present\cite{luwang:2013},
which breaks the dipolar structure of acceleration in $\theta$. The
last term is the $\theta$ symmetric part of the external velocity/
momentum source $S_{v}$. Similarly the equation for $\left\langle T_{i}\sin m\theta\right\rangle $
is obtained by multiplying the ion temperature equation by $\sin m\theta$
followed by flux surface average 
\begin{eqnarray}
 &  & \frac{\partial}{\partial t}\left(\left\langle T_{i}\sin m\theta\right\rangle -\frac{2}{3}\left\langle n\sin m\theta\right\rangle \right)-\frac{5}{3}\frac{\varepsilon_{n}}{2}\nabla_{r}\left[\left\langle T_{i}\cos(m-1)\theta\right\rangle -\left\langle T_{i}\cos(m+1)\theta\right\rangle \right]\nonumber \\
 &  & =-\sqrt{2\tau_{i}}\frac{\varepsilon_{n}}{2q}m\left\langle T_{i}\sin m\theta\right\rangle -\left\langle \left[\phi,T_{i}-\frac{2}{3}n\right]\sin m\theta\right\rangle +\left\langle S_{T}\sin m\theta\right\rangle \label{zt}
\end{eqnarray}

\selectlanguage{american}%
\noindent where the first term on the right hand side represents Landau
damping á la Hammet-Perkins\cite{hammet:1990}, the second term is
the \foreignlanguage{english}{$\sin m\theta$ weighted flux surface
averaged divergence of heat flux which has been shown to excite GAM
for $m=1$\cite{hallatschek:2001}. And the third term represents
the $\theta$ antisymmetric part of the heat source $S_{T}$.} The
above equations are supplemented by the adiabatic electron response,
which can be written as:\foreignlanguage{english}{
\begin{eqnarray}
\left\langle n\sin m\theta\right\rangle =\left\langle \phi\sin m\theta\right\rangle 
\end{eqnarray}
}These equations show that the perturbations with sine/cosine parity
and the poloidal mode number $m$ are coupled to those with the poloidal
mode numbers $m+1$, $m-1$ and cosine/sine parity. Hence we write
the equations for $m+1$ (flipping the parity):\foreignlanguage{english}{
\begin{eqnarray}
 &  & \frac{\partial}{\partial t}\left\langle n\cos(m+1)\theta\right\rangle -\frac{\partial}{\partial t}\nabla_{r}^{2}\left\langle \phi\cos(m+1)\theta\right\rangle -\frac{\varepsilon_{n}}{2}\nabla_{r}\left[\left\langle \phi\sin(m+2)\theta\right\rangle +\tau_{i}\left\langle n\sin(m+2)\theta\right\rangle \right.\nonumber \\
 &  & \left.+\tau_{i}\left\langle T_{i}\sin(m+2)\theta\right\rangle -\left\langle \phi\sin m\theta\right\rangle -\tau_{i}\left\langle n\sin m\theta\right\rangle -\tau_{i}\left\langle T_{i}\sin m\theta\right\rangle \right]+\frac{\varepsilon_{n}}{2q}(m+1)\left\langle v_{||}\sin(m+1)\theta\right\rangle \nonumber \\
 &  & =-\left\langle \vec{\nabla}\cdot\left[\phi+p_{i},\vec{\nabla}_{\perp}\phi\right]\cos(m+1)\theta\right\rangle +\left\langle S_{n}\cos(m+1)\theta\right\rangle \label{zcs1}
\end{eqnarray}
\begin{eqnarray}
 &  & \frac{\partial}{\partial t}\left\langle v_{||}\sin(m+1)\theta\right\rangle -\tau_{i}\varepsilon_{n}\nabla_{r}\left[\left\langle v_{\parallel}\cos m\theta\right\rangle -\left\langle v_{\parallel}\cos(m+2)\theta\right\rangle \right]\nonumber \\
 &  & -\frac{\varepsilon_{n}}{2q}(m+1)\left[\left(1+\tau_{i}\right)\left\langle n\cos(m+1)\theta\right\rangle +\tau_{i}\left\langle T_{i}\cos(m+1)\theta\right\rangle \right]\nonumber \\
 &  & =-\left\langle \left[\phi,v_{||}\right]\sin(m+1)\theta\right\rangle +\left\langle \left(n\nabla_{||}\left(p_{i}+n\right)\right)\sin(m+1)\theta\right\rangle +\left\langle S_{v}\sin(m+1)\theta\right\rangle \label{gv1}
\end{eqnarray}
}

\selectlanguage{english}%
\noindent 
\begin{eqnarray}
 &  & \frac{\partial}{\partial t}\left(\left\langle T_{i}\cos(m+1)\theta\right\rangle -\frac{2}{3}\left\langle n\cos(m+1)\theta\right\rangle \right)-\frac{5}{3}\frac{\varepsilon_{n}}{2}\nabla_{r}\left[\left\langle T_{i}\sin(m+2)\theta\right\rangle -\left\langle T_{i}\sin m\theta\right\rangle \right]\nonumber \\
 &  & =-\sqrt{2\tau_{i}}\frac{\varepsilon_{n}}{2q}(m+1)\left\langle T_{i}\cos(m+1)\theta\right\rangle -\left\langle \left[\phi,T_{i}-\frac{2}{3}n\right]\cos(m+1)\theta\right\rangle \\
 &  & +\left\langle S_{T}\cos(m+1)\theta\right\rangle \nonumber 
\end{eqnarray}
\begin{eqnarray}
\frac{\partial}{\partial t}\langle T_{i}\rangle-\frac{5\varepsilon_{n}}{3}\nabla_{r}\left\langle T_{i}\sin\theta\right\rangle =-\left\langle \left[\phi,T_{i}\right]\right\rangle +\left\langle S_{T_{i}}\right\rangle 
\end{eqnarray}

\noindent 
\begin{eqnarray}
\left\langle n\cos(m+1)\theta\right\rangle =\left\langle \phi\cos(m+1)\theta\right\rangle 
\end{eqnarray}

\noindent In the above equations $S_{n}$, $S_{T}$ and $S_{v}$ represents
particle, heat and momentum sources respectively. \foreignlanguage{american}{For
$m=1,3,5,...$ etc., the above set of equations yield an infinite
set of equations that describe a GAM in its full glory with all the
nonlinearities and external sources. Hence the GAM oscillations in
general appear as a result of the linear coupling of $\langle(\phi,T)\rangle$,
$\langle(n,\phi,T)\sin m\theta\rangle$, $\langle(n,\phi,T)\cos(m+1)\theta\rangle$,
$\langle v_{||}\cos m\theta\rangle$, $\langle v_{||}\sin(m+1)\theta\rangle$.}
\foreignlanguage{american}{It is apparent from the above set of equations
that not just the poloidal Reynolds stress and $\sin\theta$ component
of heat flux but all poloidal harmonics of poloidal Reynolds stress,
parallel Reynolds stess, parallel acceleration, heat flux and all
poloidal harmonics of particle, momentum and heat sources act as source/sink
of GAM excitation/damping. A detailed investigation of the role of
different poloidal harmonics of turbulent particle, momentum and heat
fluxes; and different poloidal harmonics of external sources of particle,
momentum and heat sources will be presented elsewhere. In this paper
we proceed to present the linear dynamics with poloidal mode coupling
ad infinitum. \emph{}}\\
\foreignlanguage{american}{\emph{Compact representation:-}} These
equations in general can be written as a nonlinear matrix equation
of infinite dimension in the following form 
\begin{eqnarray}
\frac{\partial G}{\partial t}=MG+N+S\label{gme}
\end{eqnarray}
where\foreignlanguage{american}{ 
\begin{equation}
G=(G_{0},G_{1},G_{2},G_{3},\cdots)^{\dagger}
\end{equation}
}$N$ is the vector of nonlinear terms and $S$ is the vector of external
source terms. The vectors in the above equation are of infinite dimensions.
\foreignlanguage{american}{$G$ is made of the sub-vectors (or orbitals)
\[
G_{0}=(\langle\phi\rangle,\langle T\rangle)
\]
\[
G_{1}=(\langle\phi\sin\theta\rangle,\langle T\sin\theta\rangle,\langle v_{||}\cos\theta\rangle)
\]
\[
G_{2}=(\langle\phi\cos2\theta\rangle,\langle T\cos2\theta\rangle,\langle v_{||}\sin2\theta\rangle)
\]
\[
G_{3}=(\langle\phi\sin3\theta\rangle,\langle T\sin3\theta\rangle,\langle v_{||}\cos3\theta\rangle)
\]
\[
\cdots\cdots etc.
\]
By defining $a=i(1+\tau_{i})\varepsilon_{n}/k_{r}$, $b=i\tau_{i}\varepsilon_{n}/k_{r}$,
$c=i5\varepsilon_{n}k_{r}/3$, $d=i(\varepsilon_{n}/2)k_{r}/(1+k_{r}^{2})$,
$e=(\varepsilon_{n}/2q)/(1+k_{r}^{2})$, $f=i\tau_{i}\varepsilon_{n}k_{r}$,
$g=\varepsilon_{n}/(2q)$ and }

\begin{figure}
\selectlanguage{english}%
\includegraphics[width=10cm,height=7cm]{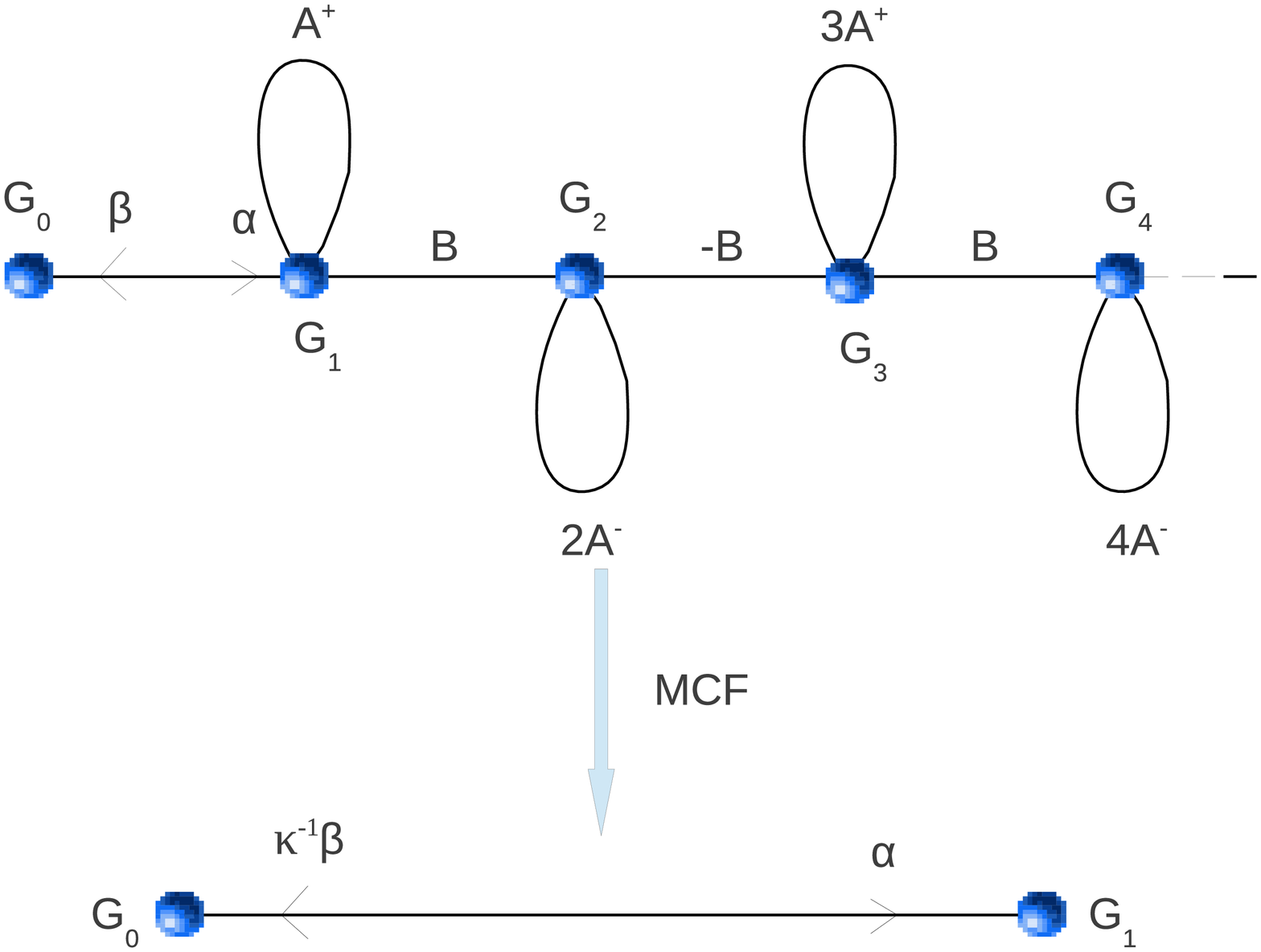}

\selectlanguage{american}%
\protect\caption{\selectlanguage{english}%
$1$\foreignlanguage{american}{d chain model of GAM }\selectlanguage{american}%
}

\label{1dc}
\end{figure}

\begin{figure}
\selectlanguage{english}%
\includegraphics[width=10cm,height=7cm]{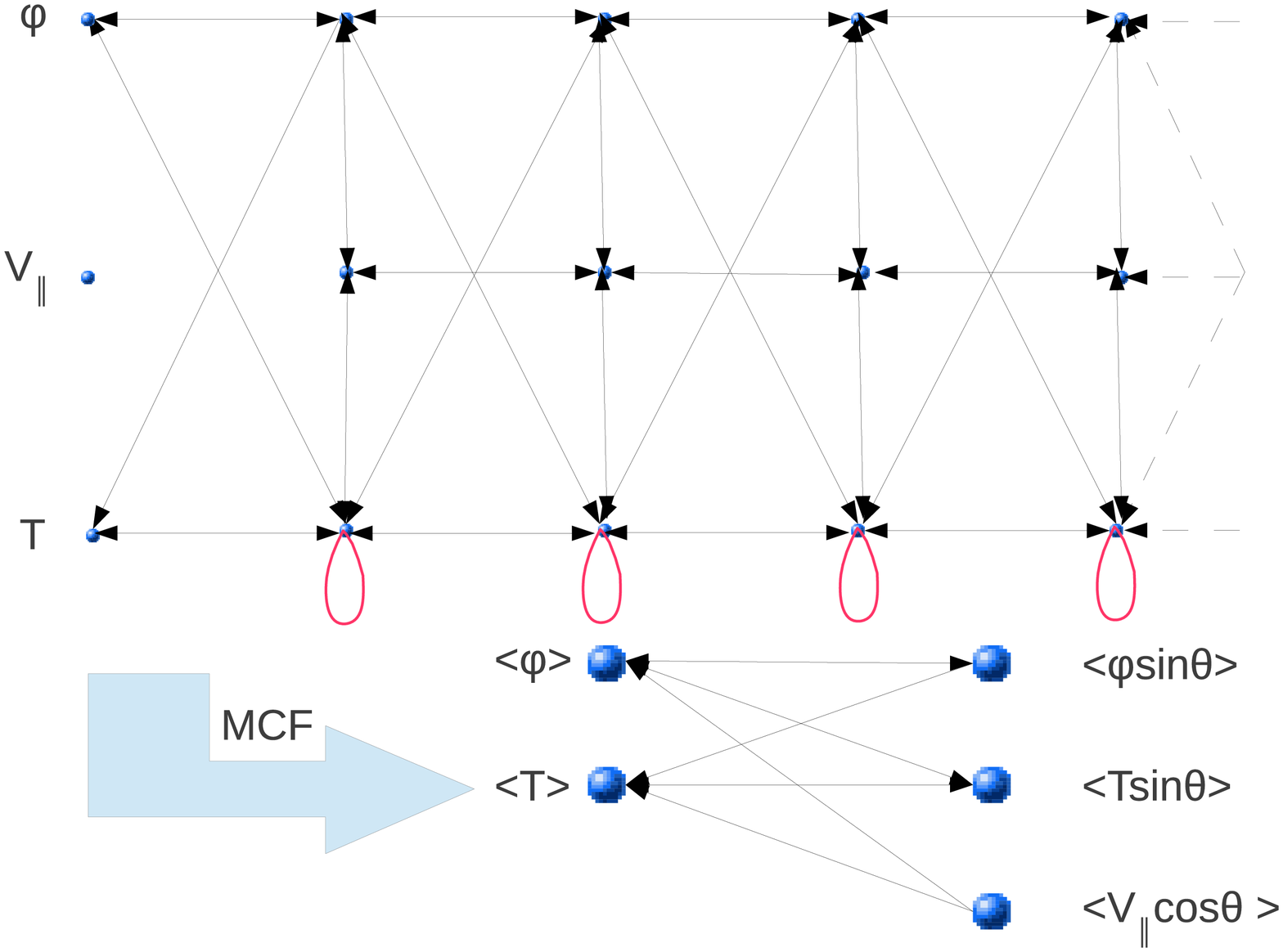}

\selectlanguage{american}%
\protect\caption{Zoom-in of FIG.\ref{1dc} }

\label{1dcz}
\end{figure}

\selectlanguage{american}%
\begin{equation}
\alpha=\left[\begin{array}{ccc}
a & b & 0\\
0 & c & 0
\end{array}\right],\beta=\left[\begin{array}{cc}
d & \tau_{i}d\\
\frac{2}{3}d & \frac{2}{3}\tau_{i}d+\frac{c}{2}\\
0 & 0
\end{array}\right]
\end{equation}
and 

\begin{multline}
\begin{array}{cc}
A^{\pm}=\left[\begin{array}{ccc}
0 & 0 & \pm e\\
0 & -\sqrt{2\tau_{i}}g & \pm\frac{2}{3}e\\
\mp(1+\tau_{i})g & \mp\tau_{i}g & 0
\end{array}\right]\\
B=\left[\begin{array}{ccc}
-(1+\tau_{i})d & -\tau_{i}d & 0\\
-\frac{2}{3}(1+\tau_{i})d & -\frac{2}{3}\tau_{i}d-\frac{c}{2} & 0\\
0 & 0 & f
\end{array}\right]
\end{array}
\end{multline}
the interaction matrix $M$ becomes a block tridiagonal matrix 
\begin{equation}
M=\left[\begin{array}{cccccccc}
0 & \alpha\\
\beta & A^{+} & B\\
 & B & 2A^{-} & -B\\
 &  & -B & 3A^{+} & B\\
 &  &  & B & 4A^{-} & -B\\
 &  &  &  & \ddots & \ddots & \ddots
\end{array}\right]
\end{equation}
Taking $G\propto e^{\lambda t}$, the equations of the orbitals become 

\begin{eqnarray}
 &  & \alpha G_{1}=\lambda G_{0}\label{G0}\\
 &  & \beta G_{0}+A_{1}G_{1}+BG_{2}=\lambda G_{1}\label{G1}\\
 &  & -(-1)^{m-1}BG_{m-1}+mA_{m}G_{m}\nonumber \\
 &  & +(-1)^{m+1}BG_{m+1}=\lambda G_{m}
\end{eqnarray}
where $m=2,3,4...etc.$, and $A_{m}=\begin{cases}
A^{+} & ;m=1,3,5,...\\
A^{-} & ;m=2,4,6,...
\end{cases}$ . A graphical representation of the above equations shows that these
orbitals form a semi-infinite chain as shown in Fig. \eqref{1dc},
where the orbitals are represented by vertices(nodes) and the couplings
are represented by the edges connecting the vertices. More detailed
couplings are shown in Fig. \eqref{1dcz} where the edges in red color
represents the self-coupling due to Landau damping.\\
\foreignlanguage{english}{\emph{Matrix Continued Fraction (MCF): }}Equations
\eqref{G1} onwards can be reduced to a single equation relating $G_{0}$
and $G_{1}$ via a matrix continued fraction(MCF) equation as follows:

\begin{eqnarray}
\beta G_{0} & = & \left\{ \left(\lambda-A^{+}\right)-B\left[\lambda-2A^{-}-B\left[\lambda-3A^{+}-B\left[\lambda-4A^{-}-\cdots\right]^{-1}B\right]^{-1}B\right]^{-1}B\right\} G_{1}\nonumber \\
 & = & \kappa(\lambda)G_{1}\label{G1R}
\end{eqnarray}
This amounts to renormalization of coupling matrix $\beta$ to $\kappa^{-1}\beta$.
Now the equations \eqref{G0}and \eqref{G1R} can be combined to get
a reduced matrix equation:

\begin{equation}
\begin{bmatrix}\lambda & -\alpha\\
-\beta & \kappa(\lambda)
\end{bmatrix}\begin{bmatrix}G_{0}\\
G_{1}
\end{bmatrix}=0
\end{equation}
for which the eigenvalues $\lambda$ can be obtained from the condition:
\[
\left|\begin{matrix}\lambda & -\alpha\\
-\beta & \kappa(\lambda)
\end{matrix}\right|=0
\]

The existence of $\kappa$ means that the infinite chain can be reduced
to a renormalized binodal chain as shown in figure\eqref{1dc}. In
this picture the minimal description of GAM is obtained when the infinite
chain is terminated at $m=1$ i.e., the binodal chain. This binodal
chain has $5$ roots. In general a chain terminated at the $m^{th}$
node has $3m+2$ roots. So the infinitely extended chain has infinite
number of roots. But the fact that the infinite chain can be reduced
to a renormalized binodal chain which again has only $5$ roots means
that only $5$ roots are physical out of the infinite number of roots
of the infinitely extended chain. 

\selectlanguage{english}%

\section{Numerical solutions and comparison with experiments\label{sec:Numerical-solutions-and}}

\subsection{Eigenvalues and amplitude spectra }

\begin{figure}[h!]
\centering{}\includegraphics[scale=0.5]{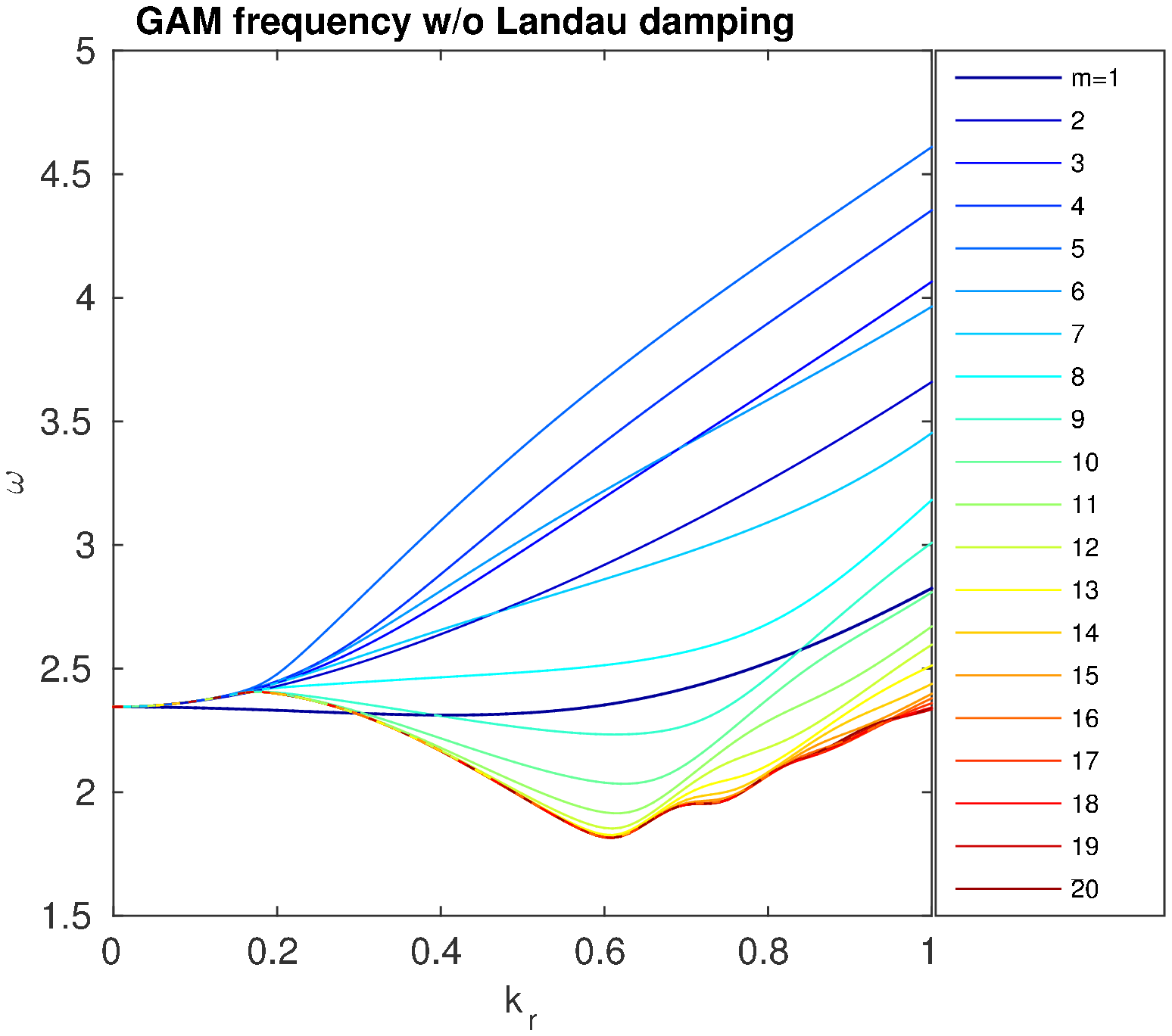}\includegraphics[width=7cm,height=7.2cm]{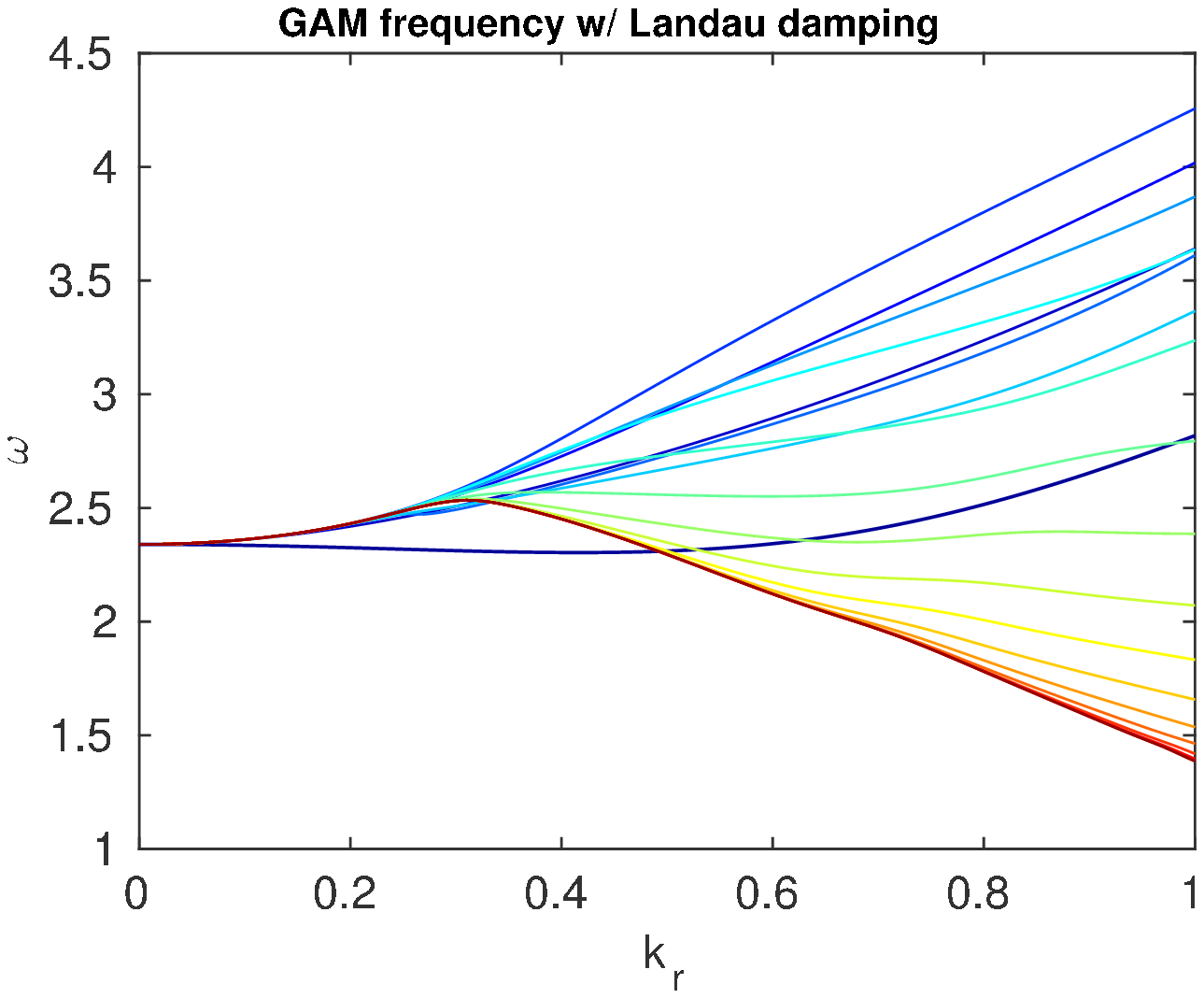}
\protect\caption{GAM dispersion convergence with increasing $m$. Left: GAM frequency
without Landau damping. Right: GAM frequency with Landau damping.
Parameters: $\tau_{i}=1$, $q=4$. }
\label{wkr}
\end{figure}

\selectlanguage{american}%
In the absence of an analytical form for the MCF $\kappa$, the matrix
$M$ is solved numerically for its eigenvalues and eigenvectors by
terminating the chain at differnt orbitals or $m$ values. \\
\emph{Root sorting algorithm: }For $k_{r}=0$ the intermodal coupling
matrix $B$ becomes a null matrix and the intermodal coupling beyond
$m>1$ vanishes. This means for closure at any $m$ the physical roots
are the ones which has the same values as the roots for $k_{r}=0$
at $m=1$ closure i.e., $\lambda(k_{r}=0,m=1)=\lambda_{i}(k_{r}=0,m=m)$.
Then the correct physical dispersion branch is followed by minimizing
$\lambda$ in $k_{r}$ i.e., minimizing $|\lambda_{i}(k_{r})-\lambda_{j}(k_{r}+\Delta k_{r})|$.
\\
\emph{Frequency:} FIG.\eqref{wkr} shows the radial wave number dispersion
of GAM frequency, with and without Landau damping, when the chain
is terminated at different nodes for the parameters $\tau_{i}=1$
and $q=4$. The dispersion curve converges at $m\ge2$ towards low
$k_{r}$ and and at $m\ge14$ towards high $k_{r}$. This clearly
means that the number of orbitals needed for convergence increases
with $k_{r}$. It is also clearly seen that Landau damping has strong
effect on the converged GAM dispersion at and beyond moderately high
$k_{r}$ values which drammatically changes the radial group propagation
properties of GAM.

 The radial group velocity can be positive, zero or negative depending
on the value of $k_{r}$ and Landau damping effect. 

\begin{figure}
\includegraphics[width=8cm,height=8.4cm]{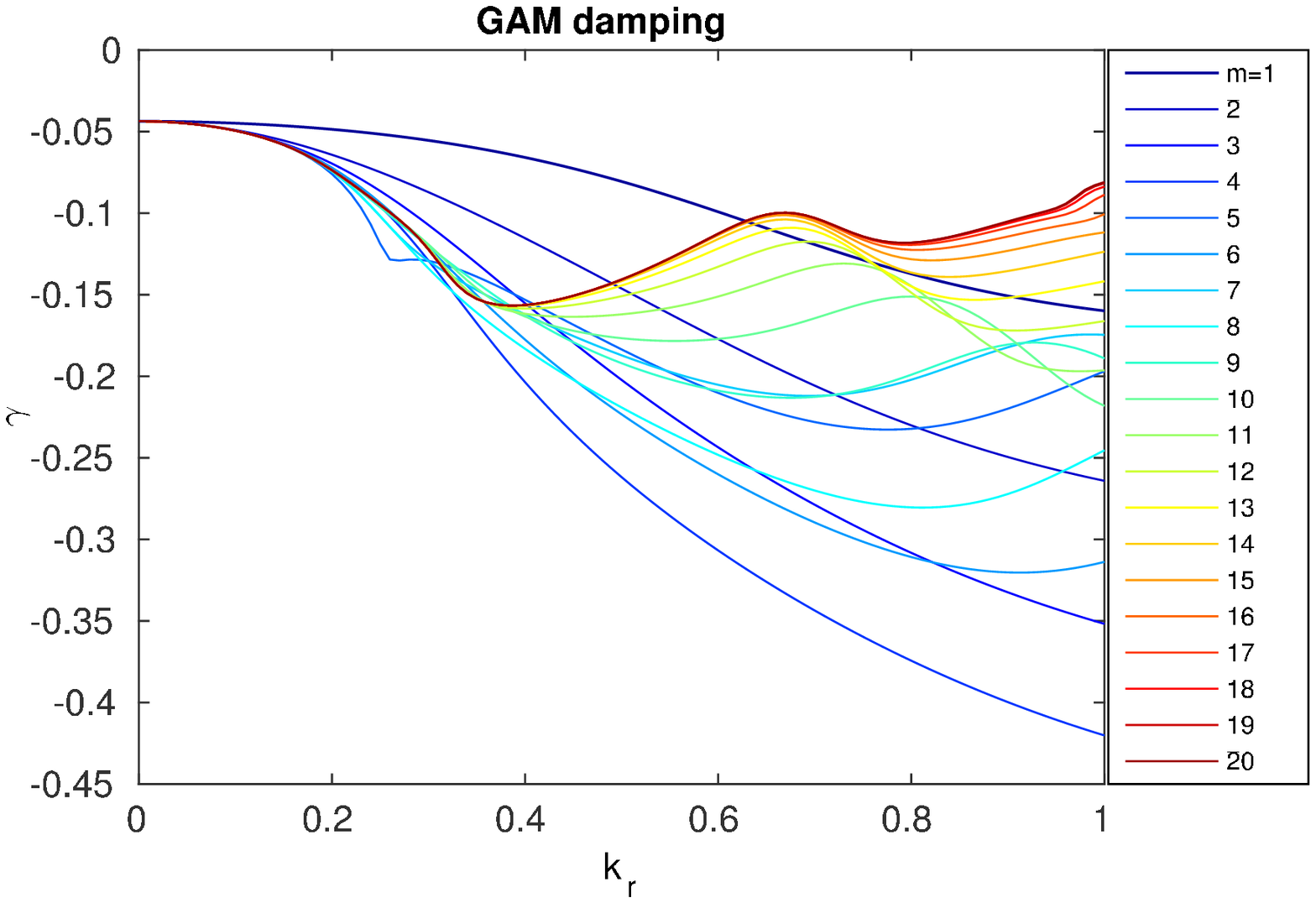}\includegraphics[width=8cm,height=8cm]{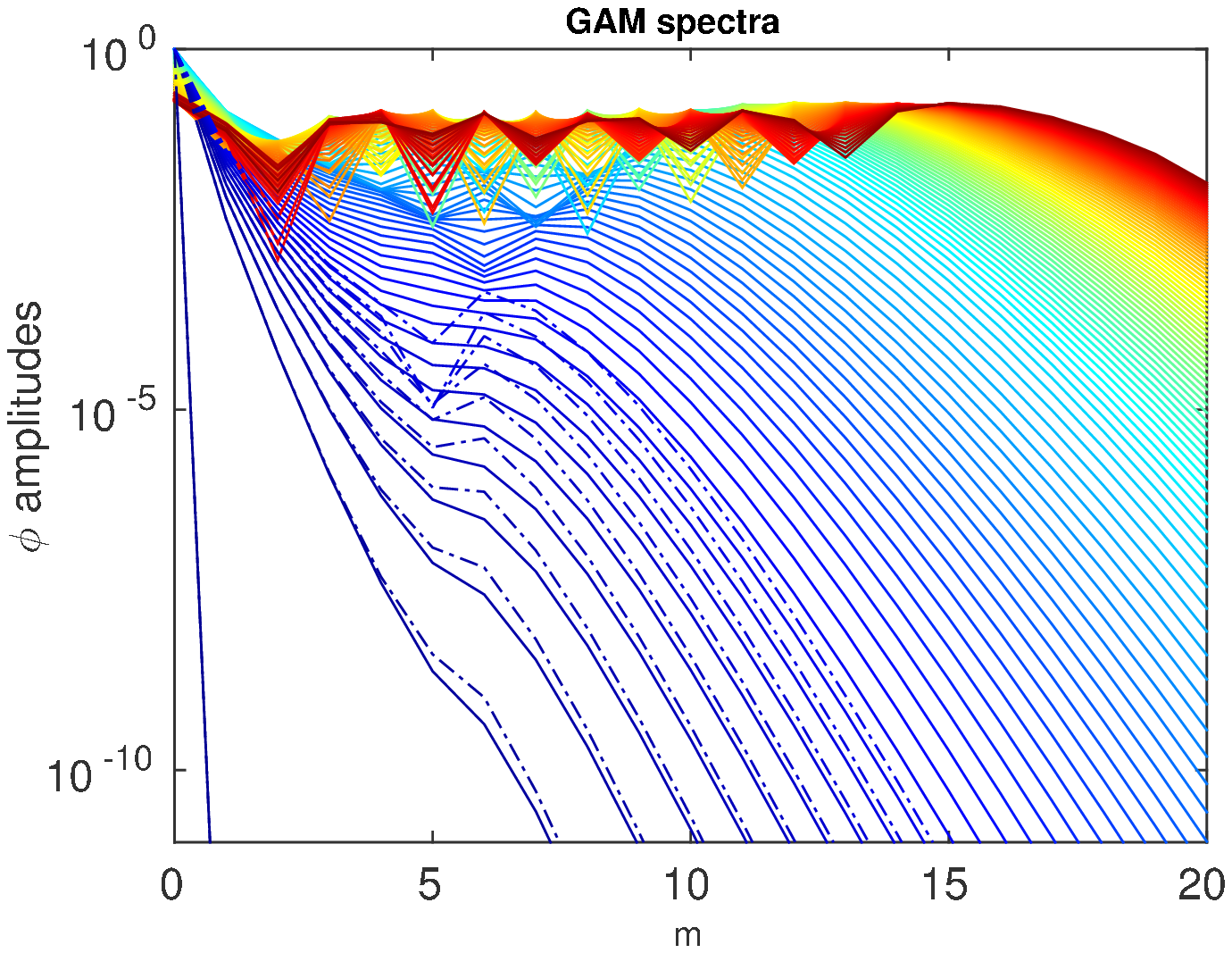}

\protect\caption{Left: GAM Landau damping. Right: GAM spectra of $\phi$ in $(m,k_{r})$.
Color coding: \foreignlanguage{english}{Blue to red $k_{r}=[0:0.01:1]$.
Dotted lines represent amplitude spectra without Landau damping.}}
\label{amp}
\end{figure}
\emph{Damping and amplitude spectra:} The damping and amplitude spectra
also confirms the same features. The left side of figure \eqref{amp}
shows that the converged damping curve is very much different from
the damping at closure $m=1$. The amplitude spectra in $m$ is plotted
for different values of $k_{r}$ in figure\eqref{amp}. The amplitude
spectra shows that the amplitude increases with $k_{r}$ at any poloidal
mode number $m$. At any $k_{r}$ the amplitude decreases rapidly
with $m$. Also the amplitude without Landau damping effect is slightly
higher than that with Landau damping. All this conveys the same consistent
lesson that poloidal mode coupling becomes important with $k_{r}$.

\subsection{Experimental comparison}

\begin{figure}
\selectlanguage{english}%
\includegraphics[scale=0.5]{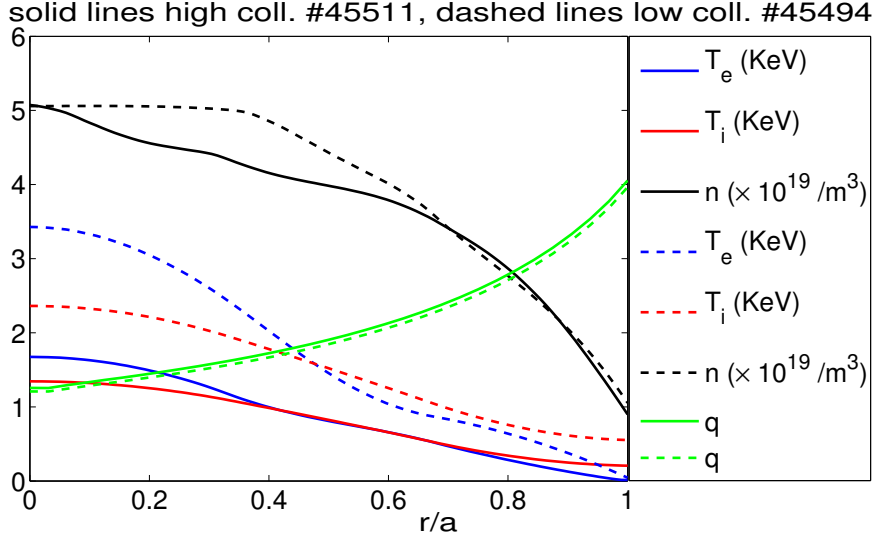}

\selectlanguage{american}%
\protect\caption{\selectlanguage{english}%
\label{fig:profiles}Equilibrium profiles of Tore Supra\selectlanguage{american}%
}
\end{figure}

\begin{figure}
\includegraphics{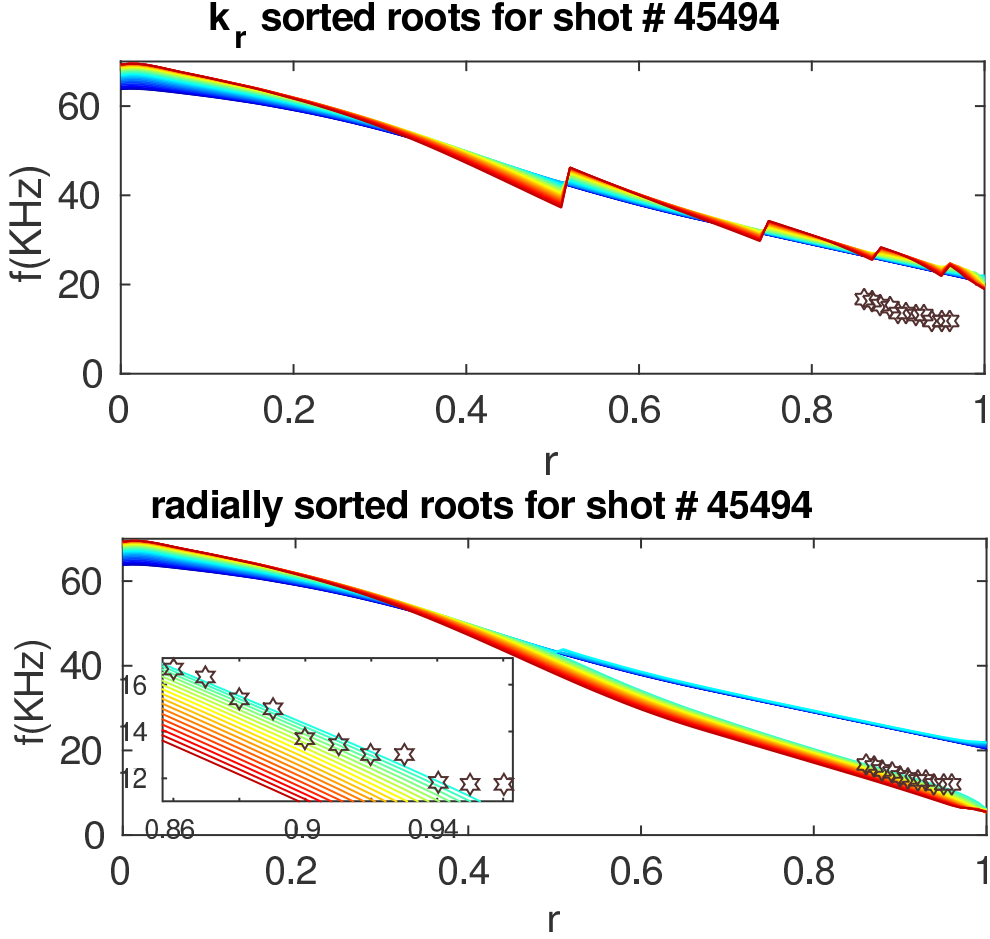}

\protect\caption{\label{fig:lowcoll}Experiment theory comparison of GAM frequency
for low collisionality shot. Top: $k_{r}$ sorted roots. Bottom: Radially
sorted roots. Color coding: blue to red $k_{r}=[0:0.01:0.3]$. Closure
at m = 15 is used for all calculations. The stars are the experimentally
measured GAM frequencies.}

\end{figure}

\begin{figure}
\includegraphics{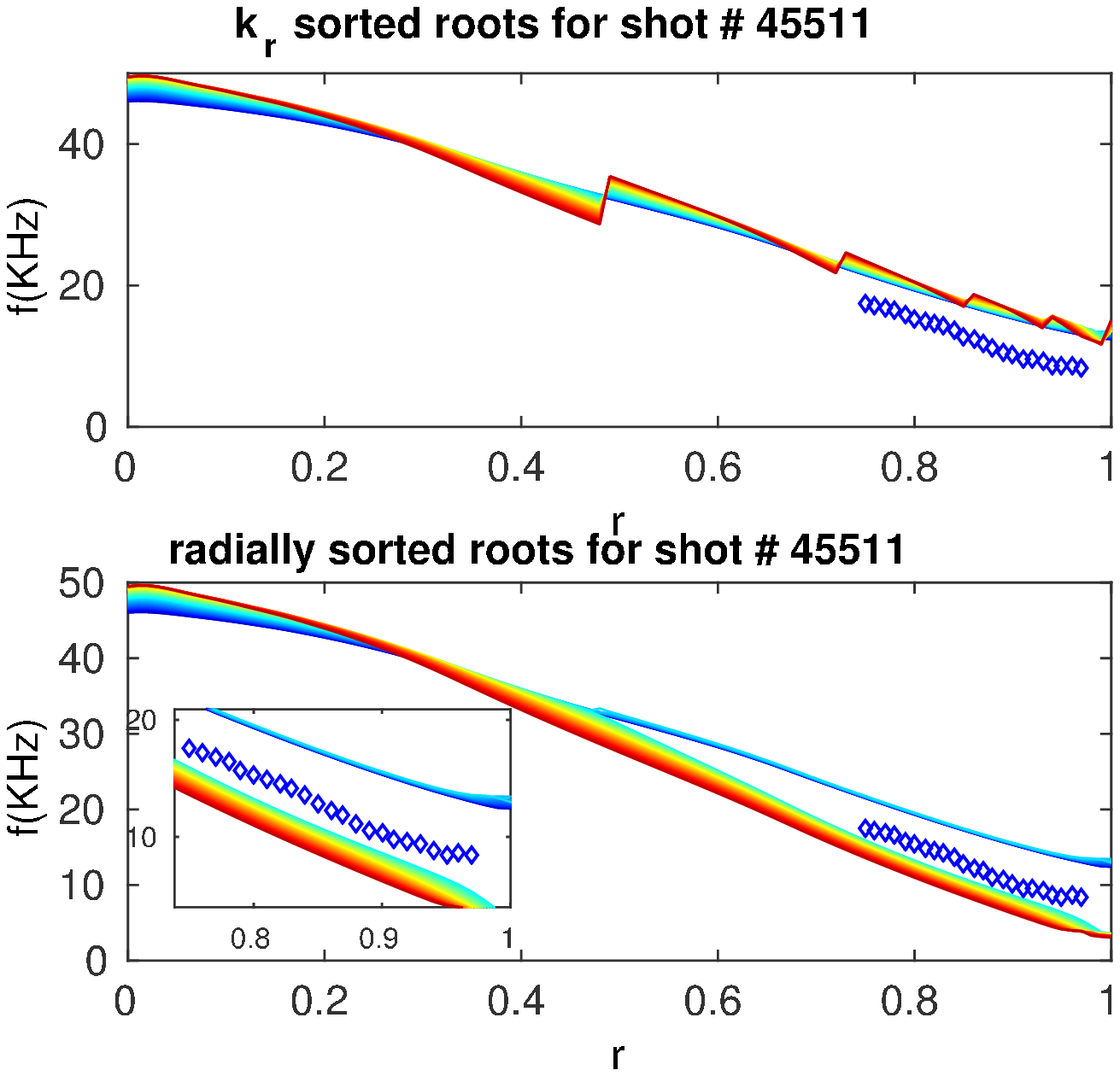}

\protect\caption{\label{fig:highcoll}Experiment theory comparison of GAM frequency
for high collisionality shot. Top: $k_{r}$ sorted roots. Bottom:
Radially sorted roots. Color coding: blue to red $k_{r}=[0:0.01:0.3]$.
Closure at m = 15 is used for all calculations. The diamonds are the
experimentally measured GAM frequencies.}
\end{figure}

\emph{Equilibrium profiles: }The equilibrium profiles for the two
different collisionality shots of Tore Supra\cite{vermare:11} are
shown in the figure\eqref{fig:profiles}. The shot number $\#45511$
is a low temperature high collisionality shot and the shot number
$\#45494$ is a high temperature low collisionality shot where densities
and the safety factors are almost same in both shots. \\
\emph{Root sorting and experiment-theory comparison: }For the given
equilibrium profiles the roots can be sorted either in radial wave
number $k_{r}$ at each radial location $r$ as described in the previous
subsection or it can be sorted in $r$ at each $k_{r}$. Radial sorting
is done by first obtaining the roots at $r=0$ by $k_{r}$ sorting.
Then $|\lambda_{i}(r)-\lambda_{j}(r+\Delta r)|$ is minimized with
respect to $r$ to follow the roots smoothly in radius. The $k_{r}$
sorted roots display discontinuous jumps in radial profile for $k_{r}>0.1$
as can be seen in the top panels of figures\eqref{fig:lowcoll} and
\eqref{fig:highcoll}. The experimentally measured frequencies are
however almost $50\%$ below the theoretically obtained $k_{r}$ sorted
roots. The radially sorted roots on the other hand, jumps in the $k_{r}$
space as can be seen in the bottom panels of the figures \eqref{fig:lowcoll}
and \eqref{fig:highcoll}. The experimental GAM frequencies overlap
with the theoretical $r$ sorted frequencies for the low collisionality
shot in the range $0.1<k_{r}<0.15$ as shown in the bottom pannel
of figure\eqref{fig:lowcoll}. However for the high collisionality
shot, the experimental values fail to overlap with the theoretical
$r$ sorted roots and falls in the spectral gap, see bottom panel
of figure\eqref{fig:highcoll}. This may be a reflection of the fact
that the effects of collision are not correctly captured by our theoretical
model. 

\selectlanguage{english}%

\section{Conclusions\label{sec:Conclusions}}

In conclusions we presented the most general picture of GAMs in tokamak
in the fluid theoretical framework starting from the Braginskii equations
applied to ITG turbulence. The unique feature of the model is that
it retains poloidal mode couplings ad infinitum which allows to study
the effects of higher poloidal mode couplings in a systematic manner.
Because of geodesic curvature each poloidal mode $m$ is coupled to
its nearest neighbors $m+1$ and $m-1$ and hence infinite number
of equations are required to describe a GAM. With appropriately defined
sub-vectors of pertubations for poloidal mode $m$, dubbed here as
orbitals, the infinite set of scalar equations become infinite set
of vector equations with $m^{th}$ orbital coupled to $m\pm1$ orbitals.
This leads to a semi-infinite $1d$ chain model of GAM with each node
represented by an orbital. This infinite chain is reduced to a bi-nodal
chain with renormalized coupling coefficients by use of a matrix continued
fraction. Since an analytical form of the matrix continued fraction
is not available, numerical methods are adopted to the dispersion,
damping and mode amplitude properties by terminating the chain at
different lengths. The main results of this paper are outlined below. 

The GAM dispersion, damping at different chain lengths show that they
converge with increasing chain lengths, the rate of convergence being
faster at low $k_{r}$ and slower at high $k_{r}$ meaning that poloidal
mode couplings are increasingly important with increasing $k_{r}$.
The poloidal coupling beyond $m=1$ vanishes at $k_{r}=0$. These
results are also supported from the linear amplitude spectra of potential
in $m$ with $k_{r}$ as a parameter. The amplitudes of perturbations
beyond $m>1$ increases with $k_{r}$ showing that high $m$ mode
couplings become more important with $k_{r}$.

The theoretical results are compared with experimental observations
of GAM frequency in Tore Supra. The frequencies sorted in $k_{r}$
show discontinuous jumps in the radial profile of frequencies and
the frequencies sorted in radial coordinate show discontinuous jumps
in $k_{r}$. It is seen that radially sorted theoretical frequencies
overlap well with the experimental frequencies in the low collisionality
shot. 

Though this is the most general theory for the description of GAM
in a fluid theoretical framework, it is not free of weaknesses and
assumptions and hence has further scope for improvements. For example
it does not consider the diamagnetic and normal curvature effects
which is justified for very low $m$ closures only. For high $m$
closure these effects become important. Such a work is in progress
and will be presented elsewhere. Hammet-Perkins closure has been used
to mimic the Landau damping in fluid model which is strictly valid
for high $m$ perturbations only. A gyrokinetic formulation of infinite
poloidal mode coupling in GAM is desirable and is left as a future
work.
\begin{acknowledgments}
The authors are thankful to Y Sarazin, and G Dif-Paradalier, T.S.
Hahm and P.H. Diamond for fruitful discussions. Thanks to A Storelli
for providing the experimental profiles in Tore Supra. This work has
been supported by funding from the ``Departement D'Enseignement-Recherche
de Physique Ecole Polytechnique'' and it has been carried out within
the framework of the EUROfusion Consortium and has received funding
from the European Union's Horizon $2020$ research and innovation
programme under grant agreement number $633053$. This work was also
partly supported by the french ``Agence National de la Recherche''
contract ANR JCJC 0403 01. The views and opinions expressed herein
do not necessarily reflect those of the European Commission. 
\end{acknowledgments}
%


\end{document}